\documentclass[runningheads]{llncs}
\usepackage[T1]{fontenc}
\usepackage{amsmath,graphicx}
\usepackage{amsfonts}
\usepackage{graphicx}
\usepackage{bm}
\begin{document}
\title{Paraformer-v2: An improved non-autoregressive transformer for noise-robust speech recognition}
\titlerunning{Paraformer-v2}
%
\author{Keyu An, Zerui Li, Zhifu Gao, Shiliang Zhang}
\authorrunning{Keyu An et al.}

%
\institute{Speech Lab, Alibaba Group, China \\
\email{{ankeyu.aky,lzr265946,zhifu.gzf,sly.zsl}@alibaba-inc.com}}
\maketitle              
\begin{abstract}
Attention-based encoder-decoder, e.g. transformer and its variants, generates the output sequence in an autoregressive (AR) manner. Despite its superior performance, AR model is computationally inefficient as its generation requires as many iterations as the output length. In this paper, we propose Paraformer-v2, an improved version of Paraformer, for fast, accurate, and noise-robust non-autoregressive speech recognition. In Paraformer-v2, we use a CTC module to extract the token embeddings, as the alternative to the continuous integrate-and-fire module in Paraformer. Extensive experiments demonstrate that Paraformer-v2 outperforms Paraformer on multiple datasets, especially on the English datasets (over 14\% improvement on WER), and is more robust in noisy environments.

\end{abstract}

\section{Introduction}
The attention-based encoder-decoder (AED) model has significantly advanced the field of Automatic Speech Recognition (ASR). However, a key limitation of the autoregressive (AR) decoder is the inherent inefficiency; each output token depends on those generated before it, leading to a decoding time directly proportional to the length of the output sequence. This linear scalability can result in slow processing, particularly troublesome for lengthy outputs.

In response, non-autoregressive transformers (NATs) have emerged as a potential solution to this bottleneck. NAT models, as exemplified by works like Mask-CTC~\cite{mask-ctc}, CASS-NAT~\cite{cass-nat}, Spike-triggered NAT~\cite{spike}, and notably Paraformer~\cite{paraformer}, revolutionize the process by concurrently generating all tokens, thereby sidestepping the sequential dependency and significantly accelerating inference.

Paraformer, in particular, stands out among NAT-based ASR models due to its success in Mandarin speech recognition. Paraformer uses a Continuous Integrate-and-Fire (CIF) based predictor to estimate token embeddings in parallel, functioning as the query input to a non-autoregressive decoder. Nevertheless, the CIF predictor confronts two principal challenges:

\textbf{Multilingual Limitations}: The CIF predictor's efficacy may wane when applied to languages such as English. Unlike Mandarin, English often employs byte-pair encoding (BPE) to tokenize text, resulting in a higher variability in token counts that the CIF struggles to accurately estimate.

\textbf{Noise Sensitivity}: The CIF mechanism exhibits reduced robustness against ambient noise, which is particularly problematic in settings prone to acoustic disturbances, like conference or meeting environments. This sensitivity can lead to a noticeable decline in recognition accuracy under suboptimal acoustic conditions.

In this paper, we introduce Paraformer-v2, a novel architecture designed to overcome the limitations observed in the original Paraformer. Central to our approach is the integration of a Connectionist Temporal Classification (CTC) module for extracting token embeddings, which performs better than CIF in language adaptability and noise-robustness. During training, the token embeddings are extracted based on the CTC alignment. During inference, we use the non-blank CTC prediction as the token embeddings to the non-autoregressive transformer decoder. Comprehensive experimental evaluations have validated the superiority of Paraformer-v2, showcasing its capability to achieve state-of-the-art performance among NAT models across multiple benchmark datasets. Specifically, our model sets new benchmarks on AISHELL-1 (a Mandarin Chinese dataset), LibriSpeech (an English corpus), and an in-house English dataset comprising 50,000 hours of speech. Notably, Paraformer-v2 rivals the performance of strong autoregressive models such as conformer AED and conformer transducers, a testament to its effectiveness. Moreover, Paraformer-v2 has demonstrated substantial improvements in noisy environments compared to Paraformer. 

\section{Methods}
We first introduce Paraformer, and then present our modifications in Paraformer-v2.
\subsection{Paraformer}
Given the speech feature ${\bf X}_{1:T}$, the encoder produces a sequence of hidden representations:
\begin{equation}
{\bf H}_{1:T} = {\rm Encoder}({\bf X}_{1:T}),
\end{equation}
and the predicted token embedding is obtained by a CIF module
\begin{equation}
{\bf E}_{1:U'} = {\rm CIF}({\bf H}_{1:T}).
\end{equation}
Specifically, CIF predicts the weights ${\bm \alpha}_{1:T} $ using
\begin{equation}
{\bm \alpha}_{1:T} = {\rm Sigmoid}({\rm Linear}({\rm Conv}({\bf H}_{1:T}))).
\end{equation}
Then, it forwardly accumulates the weights and integrates the encoder outputs until the accumulated weight reaches a given threshold $\beta_{\rm CIF}$.  It then instantly fires the integrated acoustic information for token prediction and updates the accumulated weights.  The reader is recommended to refer to CIF paper~\cite{cif} for more details.

Given the encoder hidden representations ${\bf H}_{1:T}$ and the token embedding ${\bf E}_{1:U'}$,  the final prediction is obtained by
\begin{equation}
{\bf D'}_{1:U'} = {\rm Decoder}({\bf E}_{1:U'}; {\bf H}_{1:T}; {\bf H}_{1:T}).
\end{equation}
Here ${\rm Decoder}$ is a bi-directional (non-causal) transformer decoder. 

Note that in the training stage, the weights ${\bm \alpha}_{1:T}$ are scaled by $\frac{U}{\sum_{t=1}^{T} {\alpha}_t}$ so that the predicted length $U'$ of the predicted sequence ${\bf D'}$ is equal to the length $U$ of the target sequence ${\bf Y}$.  and the model can be optimized using cross-entropy (CE) loss. A quantity loss term $\left| {\sum_{t=1}^{T} {\alpha}_t} - U \right|$ is added to the total loss to encourage the model to predict the length closer to the targets.  Thus, the total training loss is defined as
\begin{equation}
\mathcal{L} = {\rm CrossEntropyLoss}({\bf D'}_{1:U'}, {\bf Y}_{1:U}) + \left| {\sum_{t=1}^{T} {\alpha}_t} - U \right|
\end{equation}

While the CIF-based Paraformer achieved remarkable success in Mandarin speech recognition tasks, its generalizability to other languages and robustness against noisy environments were identified as key limitations.  Mandarin, being a tonal language with structured syllables comprised mainly of an initial consonant and a final vowel, presents a more predictable pattern for CIF to estimate the number of tokens in a given speech segment. Conversely, languages like English that typically employ subword tokenization methods like Byte Pair Encoding (BPE) often result in tokens that don't neatly align with phonetic or acoustic boundaries. Consequently, the CIF's performance in estimating the correct number of tokens for these languages is compromised. Another major drawback lies in the CIF's vulnerability to environmental noise. The CIF calculates weights ${\bm \alpha}_{1:T}$ independently of the actual token predictions. Noise in the input speech can produce high CIF weights, causing the model to interpret noise as meaningful tokens. 

To address the drawbacks mentioned above, we propose Paraformer-v2, as detailed in the following section.

\subsection{Paraformer-v2}
\begin{figure}
\centering
\includegraphics[width=0.25\textwidth]{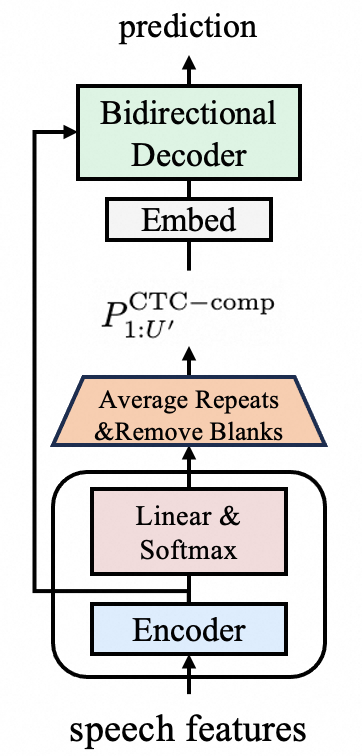}
\caption{Paraformer-v2.} \label{fig1}
\end{figure}
Different from Paraformer, we utilize a CTC module to obtain the token embedding, which proved to have better multilingual adaptability and to be more noise-robust. Specifically, we obtain the frame-wise CTC posterior $P^{\rm CTC}_{1:T}$ and CTC greedy decoding results $Y^{\rm CTC}_{1:T}$ by
\begin{equation}
P^{\rm CTC}_{1:T} = {\rm Softmax}({\rm Linear}({\bf H}_{1:T}))
\end{equation}
\begin{equation}
Y^{\rm CTC}_{1:T} = {\rm argmax}(P^{\rm CTC}_{1:T})
\end{equation}
Then we obtain the compressed CTC posterior $P^{\rm CTC-comp}_{1:U'}$ by averaging frames with the same predictions and removing blank frames in $P^{\rm CTC}_{1:T}$ according to $Y^{\rm CTC}_{1:T}$. 
\begin{equation}
P^{\rm CTC-comp}_{1:U'} = {\rm RemoveBlanks}({\rm AverageRepeats}(P^{\rm CTC}_{1:T}, Y^{\rm CTC}_{1:T}))
\end{equation}
For example, given the CTC posterior $P^{\rm CTC}_{1:5} = [p_1, p_2, p_3, p_4, p_5]$ and the greedy decoding results $Y^{\rm CTC}_{1:5} = [a, {<blank>}, b, b, c]$,  the compressed CTC posterior will be $P^{\rm CTC-comp}_{1:3} = [p_1, {\rm average}([p_3, p_4]), p_5]$.

The token embedding is defined as
\begin{equation}
{\bf E}_{1:U'} = {\rm Embed}(P^{\rm CTC-comp}_{1:U'} ).
\end{equation}
Here ${\rm Embed} \in  \mathbb{R}^{d_{\rm Dec} \times {\rm (VocabSize + 1)}} $ is a linear layer that transforms the compressed CTC posterior to the decoder input. $d_{\rm Dec}$ is the dimension of the decoder, and ${\rm VocabSize + 1}$ stands for the size of token vocabulary plus one <blank> symbol.

Similar to Paraformer, the final prediction of Paraformer-v2 is obtained by
\begin{equation}
{\bf D'}_{1:U'} = {\rm Decoder}({\bf E}_{1:U'}; {\bf H}_{1:T}; {\bf H}_{1:T}).
\end{equation}

Regarding the training process of the Paraformer-v2, a non-trivial issue is the length mismatch between the predicted token length $U'$ and the length of the ground truth  $U$, which prohibits the calculation of the cross-entropy (CE) loss. Some previous studies work around this problem by scheduling the training process to bypass the decoder optimization~\cite{spike}, or using the ground truth as the decoder input~\cite{ctc-nat} when the predicted token length is not equal to the ground truth. In contrast, we use the viterbi algorithm to find the most probable CTC alignment $\bm{\mathcal{A}}_{1:T}$ and then extract the compressed CTC posterior aligned with the target sequence based on $\bm{\mathcal{A}}_{1:T}$ :
\begin{equation}
P^{\rm CTC-comp}_{1:U'} = {\rm RemoveBlanks}({\rm AverageRepeats}(P^{\rm CTC}_{1:T}, \bm{\mathcal{A}}_{1:T}))
\end{equation}
For example, given the CTC posterior $P^{\rm CTC}_{1:5} = [p_1, p_2, p_3, p_4, p_5]$, the target sequence $Y_{1:2} = [a, b]$, and the corresponding posterior-token alignment $ \bm{\mathcal{A}}_{1:5} = [<blank>, a, <blank>, b, b]$ generated using the viterbi algorithm, the compressed CTC posterior will be $P^{\rm CTC-comp}_{1:2} = [p_2, {\rm average}([p_4, p_5])]$, which has the same length as the target sequence. The total training objective is defined as follows:
\begin{equation}
\mathcal{L} = {\rm CrossEntropyLoss}({\bf D'}_{1:U'}, {\bf Y}_{1:U}) +  {\rm CTCLoss}(P^{\rm CTC}_{1:T}, {\bf Y}_{1:U})
\end{equation}

\subsection{Experiments}
We conduct experiments on the openly available 170-hour Mandarin AISHELL-1, 960-hour English LibriSpeech, and an in-house 50000-hour English task. For all datasets, we use filterbanks as input. The input features are extracted on a window of 25ms with a 10ms shift, and then subsampled by a factor of 4 using a convolutional input layer (on AISHELL-1 and LibriSpeech), or by a factor of 6 by stacking consecutive frames (on our in-house dataset). On AISHELL-1 and LibriSpeech, we adopt a conformer encoder~\cite{conformer} and a bidirectional transformer decoder. The configurations of the encoder and decoder are shown in Table.~~\ref{config}.  On the in-house 50000-hour task, we adopt conformer or SAN-M encoder~\cite{sanm}, and a bidirectional transformer decoder. 
To gauge the models' resilience to real-world noise interference, we compile a collection of 314 authentic noise samples. An ideal noise-robust model would recognize noise inputs and produce no output (null output). We quantify the noise robustness of different models by reporting the ratio of instances where the model correctly outputs nothing among the total 314 noise samples, thereby highlighting the model's capability to discriminate between speech and non-speech signals.
\begin{table}
\caption{The configuration of Paraformer-v2 small and large.}
\label{table}
\centering
\begin{tabular}{l| c| c}
\hline
\textbf{Model} & \textbf{Paraformer-v2 (S)} & \textbf{Paraformer-v2 (L)} \\
\hline
Encoder layers & 12 & 12 \\
Encoder self-attention Dim & 256 & 512 \\
Encoder feed-forward Dim & 2048 & 2048 \\
Decoder layers & 6 & 6 \\
Decoder self-attention Dim & 256 & 512 \\
Decoder feed-forward Dim & 2048 & 2048 \\
\hline
\end{tabular}
\label{config}
\vspace{-0.5cm}
\end{table}

\begin{table}
\caption{Performance comparison on the 170 hr AISHELL-1 task.}
\centering
\begin{tabular}{l|c|c| c}
\hline
\textbf{Model} & \textbf{Size} & \textbf{ AR / NAR } & \textbf{ dev / test } \\
\hline
Conformer AED~\cite{paraformer} & 50M & AR & 4.7 / 5.2  \\
Conformer Transducer~\cite{espnet_rnnt}& 90M & AR & 4.5 / 5.0 \\
Paraformer~\cite{paraformer} & 50M & NAR & 4.6 / 5.2 \\
Paraformer, reproduced  & 50M & NAR & 4.6 / 5.1 \\
Improved CASS-NAT~\cite{improve-cass-nat} & 50M & NAR & 4.9 / 5.4 \\
CTC~\cite{scctc} & 30M & NAR & 5.7 / 6.2  \\
MaskCTC~\cite{scctc} & 30M & NAR & 5.2 / 5.7 \\
InterCTC~\cite{scctc} & 30M & NAR & 5.3 / 5.7 \\
SC-CTC~\cite{scctc} & 30M & NAR & 4.9 / 5.3  \\
\hline
Paraformer-v2 (S) & 50M & NAR & 4.5 / 4.9 \\
Paraformer-v2 (L) & 120M & NAR & \textbf{4.3} / \textbf{4.7} \\
\hline
\end{tabular}
\label{tab1}
\vspace{-0.5cm}
\end{table}

\begin{table}
\caption{Performance comparison on the 960 hr LibriSpeech task. N/A means not reported in the paper.}
\centering
\begin{tabular}{l|c|c|c }
\hline
\textbf{Model} & \textbf{Size} & \textbf{AR/NAR} & \textbf{test clean/other} \\
\hline
Conformer AED~\cite{improve-cass-nat} & 100M & AR & 3.0 / 7.0 \\
Conformer Transducer~\cite{espnet_rnnt} & 90M & AR & 2.9 / 6.8 \\
Improved CASS-NAT~\cite{improve-cass-nat} & 100M & NAR & 3.1 / 7.2 \\
Align-Refine~\cite{align-refine} & 70M & NAR  & 3.6 / 9.0 \\
Imputer~\cite{imputer} & N/A & NAR & 4.0 / 11.1 \\
A-FMLM~\cite{A-FMLM} & N/A & NAR & 6.6 / 12.2 \\
Paraformer & 50M & NAR & 6.5 / 10.7 \\
\hline
Paraformer-v2 (S) & 50M & NAR & 3.4 / 8.0 \\
Paraformer-v2 (L) & 120M & NAR & \textbf{3.0} / \textbf{6.9} \\
\hline
\end{tabular}
\label{tab2}
\vspace{-0.25cm}
\end{table}
\begin{table}
\caption{Performance comparison on the 50000 hr In-house English task. BS is short for beam search.}
\centering
\begin{tabular}{l|c|c|c|c}
\hline
\textbf{Model} & \textbf{Size} & \textbf{AR/NAR} & \textbf{BS} &  \textbf{test WER} \\
\hline
Conformer AED & 230M & AR & Y & 20.07 \\
SAN-M Transducer & 160M & AR & Y & 19.81 \\
Conformer Transducer & 180M & AR & Y & 19.11 \\
SAN-M CTC & 160M & NAR &  N & 20.16 \\
SAN-M Paraformer & 220M & NAR &  N & 22.73 \\
\hline
SAN-M Paraformer-v2 & 220M & NAR & N & 19.44 \\
Conformer Paraformer-v2 & 240M & NAR & N & 19.08 \\
\hline
\end{tabular}
\label{tab3}
\vspace{-0.25cm}
\end{table}

The results of AISHELL-1 and Librispeech are shown in the Table.~~\ref{tab1} and Table.~~\ref{tab2} respectively. It can be seen that Paraformer-v2 outperforms all NAR models in terms of character error rate (CER) and word error rate (WER). Notably, on AISHELL-1, Paraformer-v2 outperforms AR conformer and conformer transducer (with beam search) significantly. On Librispeech, Paraformer-v2 is comparable with AR conformer AED and conformer transducer with similar model sizes. 

The results of the in-house 50000-hour dataset are shown in Table.~~\ref{tab3}. SAN-M Paraformer-v2 is significantly better than the NAR SAN-M Paraformer (over 14\% improvement on WER), and comparable with the AR conformer AED and conformer transducer. Conformer Paraformer-v2 performs slightly better than SAN-M Paraformer-v2.
\begin{table}
\caption{RTFs of different models, benchmarked on Aishell-1 test set.}
\centering
\begin{tabular}{l|c|c}
\hline
\textbf{Model} & \textbf{Size} & \textbf{RTF} \\
\hline
Conformer AED, decoding beam size 5 & 50M & 0.254 \\
Conformer Paraformer & 50M &  0.011 \\
Conformer Paraformer-v2 & 50M &  \textbf{0.010} \\
\hline
\end{tabular}
\label{tab4}
\vspace{-0.25cm}
\end{table}

We use real time factor (RTF, calculated by dividing the time taken by the ASR transcription by the audio duration, benchmarked on Aishell-1 test set) to measure the inference speed, which is evaluated on a Tesla V100 GPU using batch size 1. The results are shown in Table.~~\ref{tab4}. It can be seen that Paraformer-v2 performs comparably with Paraformer in speed, and is more than 20 times faster than the AR conformer AED.
\begin{table}
\caption{Percentage of null output given noise input. The higher value indicates a better robustness against noise.}
\centering
\begin{tabular}{l|c}
\hline
\textbf{Model} &   \textbf{Percentage of null output} \\
\hline
Paraformer & 54.5 \\
Paraformer-v2 & \textbf{77.7}  \\
\hline
\end{tabular}
\label{tab5}
\vspace{-0.25cm}
\end{table}

We compare the noise robustness of Paraformer and Paraformer-v2 in Table.~~\ref{tab5}. It can be seen that Paraformer-v2 produces over 20\% less undesired output given the noise input, presumably because CTC is able to resist the noise by considering the semantic information, while CIF is more noise-sensitive as the prediction of CIF weights is semantic-independent.

\section{Conclusion}
In this paper we propose Paraformer-v2, an advancement on the original Paraformer for fast,  accurate, and noise-robust non-autoregressive speech recognition. Compared to Paraformer, Paraformer-v2 shows significantly better performance on recognition accuracy, especially on English benchmarks, and better performance on noise-robustness. Compared with the AR model such as conformer AED, Paraformer-v2 performs comparably in accuracy with over 20 times speedup.

\end{document}